\documentclass[times]{qjrms4}

\usepackage[colorlinks,bookmarksopen,bookmarksnumbered,citecolor=red,urlcolor=red]{hyperref}
\usepackage{natbib}
\usepackage{caption}
\newcommand\BibTeX{{\rmfamily B\kern-.05em \textsc{i\kern-.025em b}\kern-.08em
T\kern-.1667em\lower.7ex\hbox{E}\kern-.125emX}}

\usepackage{moreverb, amsmath}

\def\volumeyear{2018}

\begin{document}

\runningheads{K. Strommen, T.N. Palmer}{Regime dynamics and NAO predictability}

\title{Signal and noise in regime systems: a hypothesis on the predictability of the North Atlantic Oscillation}

\author{K.~Strommen\corrauth, T.N. Palmer}

\address{Oxford University, Clarendon Laboratory, AOPP, Parks Road, OX1 3PU, Oxford.}

\corraddr{Oxford University, Clarendon Laboratory, AOPP, Parks Road, OX1 3PU, Oxford. E-mail: kristian.strommen@physics.ox.ac.uk}

\begin{abstract}
Studies conducted by the UK Met Office (\cite{Dunstone2016}, \cite{Eade2014}, \cite{Scaife2014}) reported significant skill at predicting the winter NAO index with their seasonal prediction system. At the same time, a very low signal-to-noise ratio was observed, as measured using the `ratio of predictable components' (RPC) metric. We analyse both the skill and signal-to-noise ratio using a new statistical toy-model which assumes NAO predictability is driven by regime dynamics. It is shown that if the system is approximately bimodal in nature, with the model consistently underestimating the level of regime persistence each season, then both the high skill and high RPC value of the Met Office hindcasts can easily be reproduced. Underestimation of regime persistence could be attributable to any number of sources of model error, including imperfect regime structure or errors in the propagation of teleconnections. In particular, a high RPC value for a seasonal mean prediction may be expected even if the models internal level of noise is realistic.
\end{abstract}

\keywords{winter predictability, atmosphere, seasonal prediction, regimes, NAO, persistence, signal-to-noise ratio}

\maketitle

\section{Introduction} \label{sec:intro}

The North Atlantic Oscillation (NAO), being the dominant mode of variability in the North Atlantic, plays a critical role in modulating European and North-American climate, particularly in the winter months (e.g. \cite{Marshall2001}, \cite{Greatbatch2000}). For many years, seasonal prediction forecasts produced using general circulation models (GCM's) showed no significant skill at predicting the wintertime NAO, and it was widely thought that this was an inherently chaotic process with no predictability on seasonal timescales (\cite{Johansson2007}). In more recent years, this view has been challenged by various studies: see e.g. \cite{Smith2016} for an overview. In \cite{Muller2005}, some models were shown to exhibit skillful wintertime predictions of the NAO at a lead-time of 1 month, though the skill appeared limited to the shorter time-period 1987-1999; none of the models showed noteworthy skill when extended to a longer hindcast period. More recently, studies conducted using the Met Office Global Seasonal Prediction System 5 (GloSea5) displayed skillful ensemble hindcasts of the winter NAO (\cite{Scaife2014}) over a 20 year hindcast period. This skill was then corroborated by studies conducted with the UK Met Office Decadal Prediction System 3 (DePreSys3) in \cite{Dunstone2016}, which showed that this skill remained when assessed over a longer hindcast period of 35 years. Bayesian analysis conducted in \cite{Siegert2016} indicated that this skill was unlikely to be a result of chance. A multimodel study carried out in \cite{Athanasiadis2017} showed this high skill was attainable by other models as well.

At the same time, it was observed that the model seemed to have a notably low signal-to-noise ratio (\cite{Scaife2014}). Indeed, the skill of 0.62 found in \cite{Dunstone2016}, measured as the correlation between the ensemble mean of the hindcast NAO index and the NAO index as observed in re-analysis, could only be achieved by taking the mean of a large ensemble. On the other hand, the skill of the model at predicting itself (measured as the average correlation between the ensemble mean and individual members), was significantly lower, at 0.18. That is, the model NAO appears to be notably less predictable than the real NAO. This was referred to as the `signal-to-noise paradox' in \cite{Dunstone2016}. In \cite{Eade2014} and \cite{Dunstone2016} this `paradox' was encapsulated using the `ratio-of-predictable-component' ($RPC$) metric, introduced in \cite{Eade2014}. In an ideal model which captures perfectly the level of predictability of the true climate system, this metric will equal 1. An $RPC$ less than 1 indicates that the models internal NAO is too predictable, while an $RPC$ greater than 1 indicates it is not predictable enough. This metric, based on correlations, is subtly different from the perhaps more familiar `root-mean-square error divided by spread' metrics commonly used to diagnose ensemble skill relative to noise (see e.g. \cite{Fortin2014} and references within). In particular, by being insensitive to magnitudes, the $RPC$ is able to more sensitively examine situations where the signals are particularly small. The `signal-to-noise paradox' can be summarized by the observation that the $RPC$ in the DePreSys3 hindcasts is 2.31, significantly greater than 1. The fact that the estimate from the original 20 year hindcast considered in \cite{Eade2014} remained robust when extended to the 35 year hindcast considered in \cite{Dunstone2016} suggests that this estimate is not purely a result of sampling variability. While \cite{Shi2015} pointed out the sensitivity of the RPC estimate to even longer hindcast lengths, suggesting that the $RPC$ may be subject to decadal variability, it never the less left open the question as to why the $RPC$ is notably large in the last part of the century. Furthermore, given the great uncertainty inherent in estimates of the $RPC$ in situations with low model skill (see section \ref{sec:caution} and \ref{sec:corrs_markov}), a point not emphasized in \cite{Shi2015}, it is likely that more work is required to understand such decadal fluctuations.

All of this raises an obvious question: can we characterize the nature of the model skill required to produce this behaviour? In particular, is a high RPC value expected or not given the dynamics of the underlying system? The answer to this question involves the construction (or choice) of a statistical null-hypothesis, in other words a toy-model of the NAO. For example, in \cite{Scaife2014}, the work conducted in \cite{Kumar2009}, which used a simple linear shift toy-model, was cited as evidence that the observed skill was higher than what one would expect given the model's level of noise, implying that the observed RPC is spuriously high. In \cite{Siegert2016}, a `signal+noise' toy-model was used to assess the statistical significance of the observed behaviour using Bayesian techniques, concluding that both the high skill and the low signal-to-noise ratio were significant. Both these toy-models are essentially variants of the same `linear-regression paradigm'. In this paradigm of predictability, ensemble members are viewed as being normally distributed noise around an ensemble mean which detects, to some level of accuracy, a predictable signal in the \emph{magnitude} of the NAO.

Many studies have argued that the NAO exhibits regime behaviour (see e.g. \cite{Straus2007}, \cite{Woollings2010}, \cite{Corti1999}). The regime paradigm offers another potential type of predictability, namely predictability of preferred regime states. The potential importance of shifts in preferred atmospheric states for predicting the long-term climate response to anthropogenic forcing was explained in \cite{Palmer1999}, but this mechanism can be considered also on seasonal timescales. Over a given winter season, one can envisage the atmosphere, on each day, falling into a particular regime state. If the models atmosphere has a preference towards spending more days in regimes that induce positive NAO indices (for example due to some external forcing), one would expect the seasonal mean to be positive, and similarly for negative states. If the preferred model state coincides with the preferred state in reality, this will induce positive correlation. In this way, one can envisage that in a given season it may be possible to have predictability of the \emph{sign} of the NAO index which is independent of any predictability of the \emph{magnitude}. Indeed, it may be that in some seasons, the sign is essentially the \emph{only} thing predictable.

In this paper we will analyse a new toy-model aimed at capturing such regime behaviour. In this toy-model, the NAO is explicitly represented as an idealized 2-state Markov chain, with seasonally varying transition probabilities. Error is introduced into ensemble members by modulating the extent to which they are able to capture these probabilities. Specifically, we will assume that, ignoring noise from internal variability, the ensemble members all underestimate the persistence probabilities of both regimes. We show that this toy-model can easily capture the observed high-$RPC$ behaviour, even when the internal level of noise in the model perfectly matches that of observations: the apparently excessive noise on seasonal timescales is entirely a consequence of model error on daily timescales. Concretely, the combination of high ensemble mean correlations and high RPC values is to be expected in this toy-model due to the weak persistence alone, and as such, may be attributable to any number of model errors, including e.g. imperfect regime structure or errors in the propagation of teleconnections. In particular, low signal-to-noise ratios can be expected due to model bias even when the models internal variability (i.e. noise) is perfectly realistic.

Finally, we compare the RPC metric to another common way to measure if a model is under/over-confident, namely the ratio of the root-mean-square error of the ensemble mean prediction with the ensemble spread. It is shown that, unless the GCM is statistically perfect, these two metrics in general may give entirely different answers.

Because the skill observed in \cite{Dunstone2016} with DePreSys3 is so high over a long 35-year window, we will be focusing on replicating the explicit numbers from that study in this paper. While the results will therefore be aiming to reproduce this particular example of the `signal-to-noise paradox', we note that we believe the model presented could equally well be fitted to data from other GCMs with comparable results.

\section{Methodology and definitions}
\subsection{Definition of relevant metrics}
\label{sec:rpc_def}

We will primarily be interested in three different metrics assessing model skill, all closely related. We define $Corr(EnsMean, Obs)$ to be the Pearson correlation coefficient between the ensemble mean of some forecast and observations over a given time-period. We will refer to this informally as the \emph{actual predictability}. On the other hand, we define $Corr(EnsMean, Mem)$ to be the average correlation between the ensemble mean and individual members of the ensemble. This is sometimes referred to as the `potential predictability', but we will avoid this term as it is misleading. Indeed, as we shall see, this quantity can often be significantly lower than the actual predictability, and therefore does not provide a well-defined upper bound on the level of predictability.\footnote{This observation was also noted in \cite{Kumar2014}.} nstead, we will refer to this quantity as the \emph{model predictability} of the system, as it measures the models ability to predict itself.

The third metric of interest is the `ratio of predictable components', $RPC$, defined in \cite{Eade2014}. One first defines the notion of the `predictable component' of a given system as the square root of the total fraction of variance that can be predicted in that system. When applied to observations we obtain $PC(Obs)$, and when applied to a model we obtain $PC(Mod)$. The $RPC$ is then by definition
\begin{equation}
RPC_{true} = PC(Obs)/PC(Mod)
\end{equation}
In \cite{Eade2014} they provide an estimate of $RPC$ amenable to computation via the following inequality:
\begin{equation}
RPC_{true} = \frac{PC(Obs)}{PC(Mod)} \geq \frac{\sqrt{Corr(EnsMean, Obs)^2}}{\sqrt{\sigma_{sig}^2 /  \sigma_{tot}^2}} 
\end{equation}
where $\sigma_{sig}^2$ is the ensemble mean variance and $\sigma_{tot}^2$ is the average variance of individual ensemble members. We define the RPC estimate to be this last term:
\begin{equation}
RPC_{est} = \frac{Corr(EnsMean, Obs)}{\sqrt{\sigma_{sig}^2 /  \sigma_{tot}^2}} 
\end{equation}
It is important to note that this lower bound has been obtained by choosing a positive square root, and so is always positive, even if the correlation between ensemble mean and observations could, in principle, be negative. Thus our estimate of the $RPC$ is in fact the absolute value of this last term.\footnote{Note that in \cite{Siegert2016} this estimate (without the absolute value) is taken as the \emph{definition} of $RPC$. We will stick with the original definition in this paper.}

For a model to be `statistically perfect', one firstly requires that different ensemble members are statistically indistinguishable from each other, and therefore that they share the same variance: $\sigma_{tot} = \sigma_i$ for all $i$, where $\sigma_i$ is the standard deviation of ensemble member $i$. Given this assumption, it is easy to show that the expected value of the denominator of the $RPC$ estimate above is exactly equal to the model predictability, and so $RPC_{est}$ equals the ratio of actual to model predictability. Secondly, for a `perfect' model, observations are required to be statistically indistinguishable from individual members (a property also known as `exchangability'). Under this assumption, it is easy to show that, in expectation, the internal and actual predictability will coincide. 
If we define `statistical perfection' to be met when both these conditions are met, it follows that for a perfect model, the $RPC$ will equal 1. An $RPC >1$ corresponds to the model being more skillful at predicting observations than itself,  with the opposite being true when $RPC$ $<1$. 

In practice, in the cases considered in our analysis, the assumption that different ensemble members share the same variance will not typically be true due to sampling variability and the fact that the ensembles considered will only have 40 members. Therefore, when computing it we always use the explicit estimate. However, equating $RPC$ with the ratio of actual to model predictability is useful for guiding intuition and interpretation.

\subsection{A cautionary note on computing the RPC}
\label{sec:caution}

The estimate $RPC_{est}$ is obtained by using the ensemble mean as a proxy for reality, so one can compute $PC(Obs)$. If the model is very skillful, then even a small ensemble will produce an ensemble mean which is a good approximation to reality. However, in situations where the inherent model skill is low, this will no longer be the case. In fact, given a fixed ensemble size, the estimate breaks down dramatically as model skill decreases.

Indeed, if the model has no skill whatsoever, then both actual and model predictability are expected to be 0. In particular, $PC(Mod) = 0$ and the ensemble members and observations are all uncorrelated. However, assuming that the true system has \emph{some} predictable component, $PC(Obs) > 0$, and so $RPC_{true} = \infty$. On the other hand, if we assume that in this `no-skill' situation, all the ensemble members and reality are behaving as independent normal distributions, we can approximate all the terms in equation (3). If we assume that each individual ensemble member follows a distribution with variance $\sigma_{m}^2$, and that there are enough ensemble members to make $\sigma_{tot} = \sigma_{m}$ a decent approximation, then the denominator of (3) will be $1/\sqrt{N}$. To estimate the numerator, first note that the distribution of correlation coefficients between independent normal distributions follows a Student's T-distribution, which, given a sufficiently large sample size $K$ can be approximated as normal with mean 0 and variance $1/\sqrt{K}$. Since we are taking absolute values, we also need to note that if $Y$ is a normal distribution with mean 0 and variance $\sigma^2$, then $|Y|$ has mean $\sigma\sqrt{2/\pi}$. Putting all this together, we find that when the model has no skill,
\begin{equation}
RPC_{est} \approx \sqrt{2N/\pi K}
\end{equation}
where $N$ is the ensemble size and $K$ is the length of the timeseries being correlated.\footnote{We will see later that this approximation is very good in the situation we are in.} For example, with $N=40$ and $K=35$, as in \cite{Dunstone2016}, this is approximately $0.85$. If one does not take absolute values in $RPC_{est}$, the expected $RPC$ estimate can similarly be shown to be exactly 0. So if the model has little to no skill, and the ensemble size $N$ has been fixed, the true $RPC$ will diverge to $\infty$, while the estimate could be close to 0 or 1 depending on how it is computed! On the other hand, if we leave model skill fixed (e.g. by fixing a model cycle) then since the expression (4) diverges to $\infty$ as $N$ grows, $RPC_{est}$ can be made into an arbitrarily better estimate of $RPC_{true}$ by increasing the ensemble size.

The moral of the story is that:
\begin{itemize}
\item[(1)] It is important to be clear about exactly how one has computed the $RPC$ estimate, in particular whether absolute values have been taken or not.
\item[(2)] In situations where the model skill or ensemble size is small, great caution must be taken when interpreting estimates of $RPC_{true}$. Depending on the method used, an $RPC_{est}$ close to 0 or 1 may be indicative of nothing more than the fact that the actual predictability is low, and cannot be used to infer anything about the true $RPC$.
\end{itemize}

\noindent This is of particular importance when e.g. computing $RPC$ over time-periods where the model exhibits little skill, where a typical, modern ensemble size of around 40-50 may be insufficient to robustly estimate it.

\subsection{Data}
\label{sec:data}

The DePreSys3 ensemble data was kindly provided to us by Nick Dunstone, and is what we will be comparing our toy-model against. When the observational NAO index is referred to, we are using the same observational dataset considered in \cite{Dunstone2016}, namely HadSLP (see \cite{Allan2006}). This data was also generously provided to us by Nick Dunstone. In this way, we can faithfully compare our toy-models to the results of \cite{Dunstone2016}. In their paper, the NAO was calculated as the difference in mean sea level pressure averaged over boxes around Iceland and the Azores.

The determination of various regime parameters in section \ref{sec:markov} was done using the reanalysis product ERA-Interim (\cite{Dee2011}). As an example of models lack of persistence, we used data from GloSea5\footnote{Note that this is the same model as used in the DePreSys3 system.} (see \cite{Maclachlan}), available freely via the Copernicus Project (\cite{Raoult2017}). This dataset consisted of 21 ensemble members covering the period 1993-11-01 to 2016-03-01, thereby comprising 483 years of model data.

\section{The Markov model}
\label{sec:regimes}

\subsection{The NAO index as a 2-state Markov chain}
\label{sec:markov}

The signal+noise model of \cite{Siegert2016} assumed that the seasonally varying probability distribution function (pdf) of possible NAO indices is normally distributed with a mean which is shifted linearly from season to season according to a predictable scaling factor. Model skill is accordingly induced by the extent to which the model correctly detects this linear shift. In this section we introduce our new Markov model, aiming to capture the predictability expected in a more non-linear, regime based system. We will aim to show that such a model provides another hypothetical explanation for the `signal-to-noise paradox'.

From a regime perspective, a season is often thought of as being built up of daily events (e.g. \citep{Dawson2012}), a convention we will follow. A DJF season thereby consists of approximately 90 days, where for each day, the atmosphere resides in one of the available regimes. In this way, a 90 day mean is generated by sampling a Markov process over 90 steps, and as such is determined both by the magnitude of individual events and the matrix of transition probabilities. If the discrepancy between the magnitudes of events in different states is small, the long-term mean will be determined almost uniquely by the transition probabilities alone.

\begin{figure}[p]
	\centering
    \includegraphics[width=0.8\textwidth]{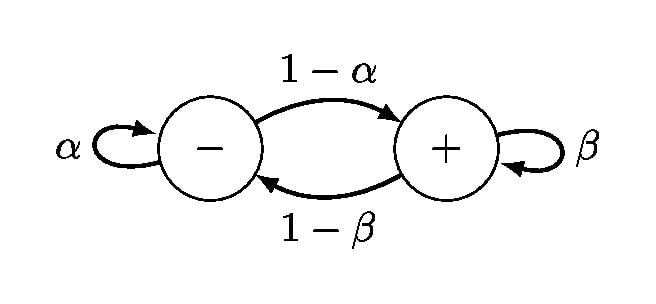}
	\caption{Illustration of a 2-state Markov chain with persistence probabilities $\alpha$ and $\beta$.}
	\label{fig:markovchain}
\end{figure}

This can be made explicit in simple situations. Figure \ref{fig:markovchain} shows a schemata for a basic, 2-state Markov chain with transition probabilities $\alpha$ and $\beta$ (ranging between 0 and 1) and two states labelled - and +, corresponding to the NAO- and NAO+. Suppose that each timestep in this chain corresponds to one day, and that each day the atmosphere ends up in the minus state, the value of the NAO index on that day is a random draw from the probability distribution $X_{-}$. Conversely, on days where it ends up in the plus state, the NAO index is a random draw from $X_{+}$. Denote the pdf's of both as $\rho_{+}(x)$ and $\rho_{-}(x)$ respectively.

Let $\pi_{-}$ and $\pi_{+}$ denote the long-term average proportion of time spent in state $-$ and $+$ respectively. It is an exercise in basic Markov theory (see Appendix A for a sketch of the derivation) to show that 
\begin{eqnarray}
\pi_{-} = (1-\beta) / (2 - \alpha - \beta) \nonumber \\
\pi_{+} = (1-\alpha) / (2 - \alpha - \beta) \nonumber
\end{eqnarray}
and hence the preferred state is determined by the relative values of $\alpha$ and $\beta$, with no preferred state if and only if $\alpha=\beta$.

If $Y = Y(\alpha, \beta)$ represents the distribution of N-day means obtained from this process, then for sufficiently large values of $N$, we have
\begin{eqnarray}
Y & \sim & \frac{1}{N} \left( \sum_{N(+)} X_{+} + \sum_{N(-)} X_{-} \right) \\
& \sim & \frac{1}{N(+)} \sum_{N(+)} \pi_{+} X_{+} + \frac{1}{N(-)} \sum_{N(-)} \pi_{-} X_{-}
\end{eqnarray}
where $N(+) \approx N\pi_{+}$ is the expected number of timesteps spent in the $+$ state and $N(-) \approx N\pi_{-}$ the expected number of timesteps spent in the $-$ state. By the Central Limit Theorem, and the fact that the sum of normal distributions is normal, $Y$ will be approximately normally distributed with expected value
\begin{equation}
E(Y) = \pi_{+}E(X_{+}) + \pi_{-}E(X_{-})
\end{equation}
and variance
\begin{equation}
Var(Y) = \frac{\pi_{+}Var(X_{+}) + \pi_{-}Var(X_{-})}{N}
\end{equation}

Let us further assume that  $X_{+}$ is strictly positive, and that $\rho_{-}(-x) = \rho_{+}(x)$ for all $x>0$; in other words, NAO+ days always have positive index, and NAO- days always have negative index, with the distribution of each being equal up to a sign. If $\beta > \alpha$, then $\pi_{+} > \pi_{-}$. The NAO+ is now the preferred state, and the expected mean and variance of $Y$ is 
\begin{equation}
E(Y) = (\pi_{+} - \pi_{-})E(X_{+})
\end{equation}
and
\begin{equation}
Var(Y) = Var(X_{+})/N
\end{equation}
From equation (5) we can further deduce that, for sufficiently large $N$, the $N(+)$ negative contributions from $X_{-}$ will approximately cancel an equal number of positive contributions from $X_{+}$, leaving on average $N(+) - N(-)$ strictly positive contributions. Therefore, $Y$ is expected to be strictly positive. The situation is opposite for NAO- years when $Y$ is expected to be strictly negative.

In a seasonal mean, $N=90$. If $\alpha$ and $\beta$ are close to each other, this may not be a sufficient number of days to witness the above expected trend, even with a statistically perfect model. In particular, the pdf of seasonal means may not end up being strictly positive or negative. This would correspond to a year with a low level of predictability. If the probabilities are more well separated on the other hand, one would be able to detect the impact on the mean robustly over a short period of time, corresponding to a year with a high level of predictability. This predictability would be induced by seasonal deviations in the persistence/transition probabilities from their climatological means, and the models ability to detect these. Such deviations could be driven by some specific external forcing (e.g. tropical sea surface temperatures, the stratosphere, sea ice extent, etc. as suggested in \cite{Dunstone2016}), internal dynamics (e.g. transient baroclinic eddy feedback), or a complex combination of the two.

In the papers \cite{Dawson2012}, \cite{Dawson2015}, clustering algorithms are used to argue for the existence of four distinct regimes, commonly titled NAO+, NAO-, Blocking and Atlantic Ridge. However, other studies suggest there may be three (\cite{Woollings2010}), or even only two regimes (\cite{Woollings2008}), suggesting the picture remains uncertain. For the sake of simplicity, we will view the North Atlantic as a 2-state system, in line with \cite{Woollings2008}, where positive NAO events correspond to more consistently zonal, unblocked jet, while the negative NAO corresponds to more consistently blocked jet. In terms of the four regime picture of \cite{Dawson2012}, this amounts to amalgamating the NAO-, Blocking and Atlantic ridge regimes into one big regime representing blocked flow, and leaving the NAO+ as representing zonal flow. 

Before testing this model in the next section, we point out that this should be seen as a `proof-of-concept' model. It seems unlikely that the NAO could be easily captured in its entirety by such a simple 2-state process, but while one could potentially extend the analysis to a multi-state system, we will see that with two states one already captures many features of the observed NAO predictability. An extension of this framework to four states will be considered in future work.

\subsection{Potential sources of model error}
\label{sec:error_sources}

Using, as explained above, the 2-state Markov chain as a toy-model for the NAO index, any GCM aiming to predict the NAO needs to accurately predict changes in the transition/persistence probabilities in each season in order to capture the predictable component. In reality, models are known to suffer from systematic biases, and so errors are expected to be present in the Markov model. We propose four different potential sources of model error that may be expected to interact with the regime structure:

\begin{itemize}
\item[(1)] There is too much inherent stochasticity in the system. Imagining the regime states as potential wells, this corresponds to a model where if the atmosphere ends up in a given well, it will quickly be pushed out of it due to the high level of noise in the system, systematically warping estimates of transition probabilities.
\item[(2)] The regime structure in the model is too weak, or suffers from persistent biases. In this setting, there is no excess stochasticity in the system, but the potential wells themselves may for example be too shallow, meaning that even realistic levels of noise is sufficient to push the atmosphere out of a given regime too quickly.
\item[(3)] If we assume that variations in transition probabilities from year to year are driven to a large extent by specific sources of external forcing (i.e. global dynamics), then it may be that the model is simply unable to correctly propagate this forcing consistently. In this way, even if the model has perfect regime structure \emph{and} realistic levels of noise, it may still fail to predict the correct probabilities. 
\item[(4)] If we instead assume that internal dynamics are the key driver to changes in transition probabilities, it may be that even tiny errors in initial conditions are sufficient to produce ensemble members that fail to capture the observed transition probabilities, even if the model is otherwise very skillful.
\end{itemize}

If one assumes that the North Atlantic does have regimes, then any error would necessarily filter through the regimes in a manner not obvious a priori. Therefore, while only error (2) in the above list specifically pertains to deficiencies in the regimes themselves, the other errors may still be important to understand in the context of our statistical model. However, precisely because only error (2) is explicitly related to regime deficiencies, we will focus on this possibility in the remainder of the paper.

That climate models tend to have persistent biases when it comes to their representation of North Atlantic regime structure has been observed in e.g. \cite{Dawson2012}. Indeed, in re-analysis, persistence is typically more common than transition, with persistence probabilities of all North Atlantic regimes being greater than one half. However, models typically do not obtain similar levels of persistence, struggling e.g. to achieve long-lived blocking events (\cite{Scaife2010}, \cite{Davini2016}). This implies that a typical model error is to underestimate persistence probabilities. This would be consistent with both (1) and (2) above, as both might be expected to reduce model persistence.

\begin{figure}[p]
	\centering
    \includegraphics[width=0.8\textwidth]{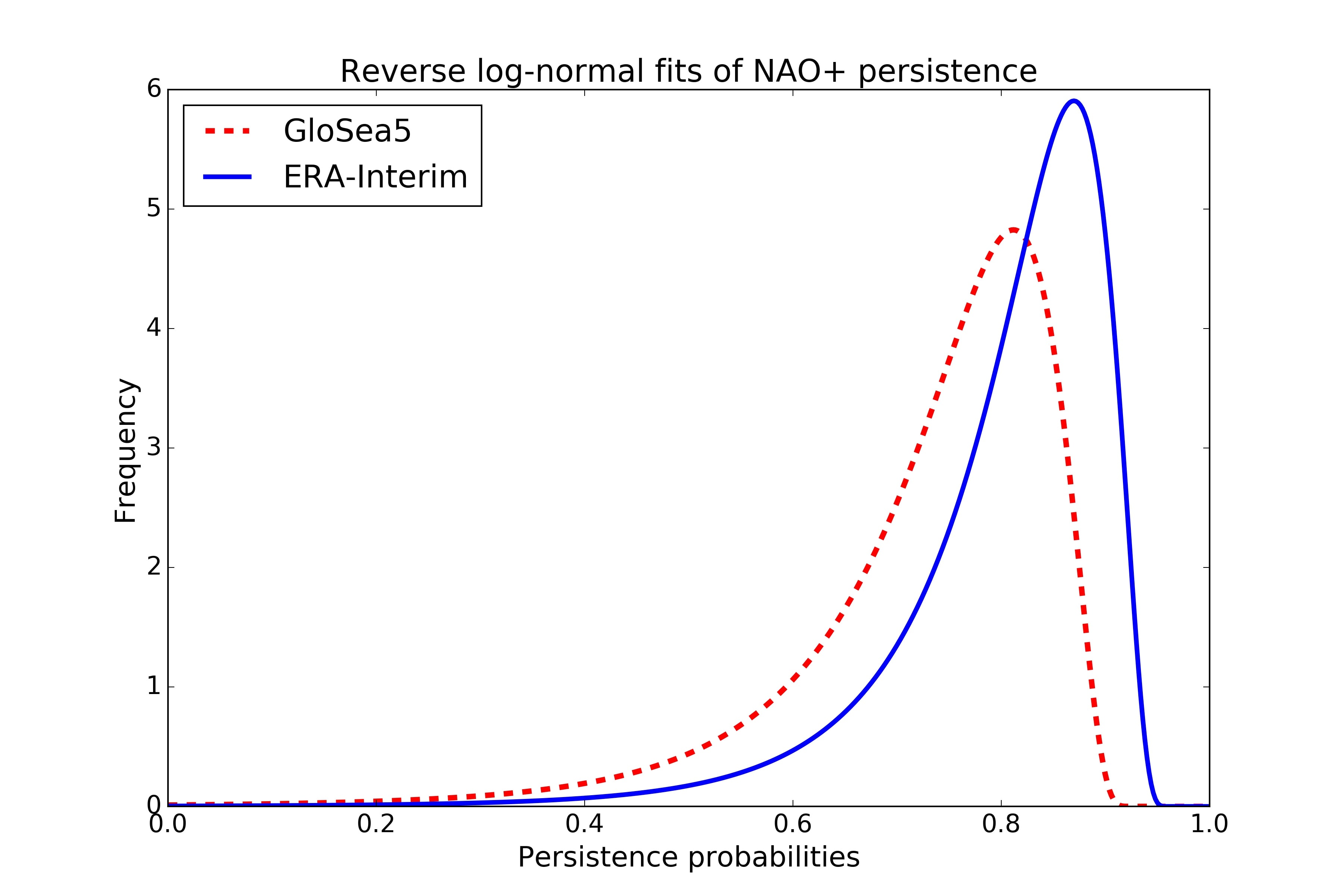}
	\caption{Reverse log-normal distributions fitted to estimates of NAO+ persistence probabilities for ERA-Interim (blue solid) and GloSea5 (red dotted). }
	\label{fig:persistence_pdfs}
\end{figure}

As a concrete example of this, we compare persistence of the NAO+ regime in ERA-Interim to that in the GloSea5 model: we remind the reader that this is the same underlying model as DePreSys3 (see section \ref{sec:data}). To do so, the method detailed in \cite{Dawson2012} was used to compute regimes in both datasets using K-means clustering. The clustering is performed on daily, detrended anomalies of geopotential height at 500 hPa over the Euro-Atlantic region in the DJF season; the dimensionality is first reduced by projecting re-analysis and model data onto the 10 leading empirical orthogonal functions of ERA-Interim. Four significant clusters are identified, corresponding to the usual four regimes, and each day of re-analysis and model data is then sorted into one of the four regimes. From this, persistence probabilities can easily be computed.\footnote{This cluster analysis was carried out in collaboration with I. Mavilia (CNR); a detailed analysis of the full results will appear in a forthcoming paper.} Visual inspection suggests that histograms of persistence probabilities are distributed according to a reverse log-normal distribution (i.e. a distribution $X$ such that $-X$ is log-normal). Figure \ref{fig:persistence_pdfs} shows the result of fitting such a distribution to the estimates of persistence probabilities of the NAO+ regime.\footnote{This fitting was carried out using the Python package `scipy', which has a standard routine for fitting log-normal distributions. By applying this to the transition probabilities, we obtain a reverse log-normal distribution of persistence probabilities.} It can be clearly seen that GloSea5 underestimates persistence, with too many short-lived NAO+ events. Similar behaviour is seen for the other regimes (not pictured). We note that this was already observed independently in \cite{Matsueda2018}, which shows examples of similar errors in other models as well. While this observation suggests weak persistence is a notable model bias, this could in principle also be attributable to any of the four suggested sources of error above, and therefore may be e.g. due to a weak signal rather than weak regimes, as suggested in \cite{Siegert2016}.

In the next section we will formulate a method to test the impact of this weak persistence on the Markov model, by systematically relaxing the seasonal persistence probabilities of the $GCM$ back towards $1/2$. The idea is simple. For each DJF season to be simulated in our hindcast, we randomly draw two persistence probabilities $p_{+}$ and $p_{-}$ for the two regimes: these represent the true probabilities that reality follow. We then generate a DJF mean for reality by taking a 90-day random walk generated with these probabilities, where the distributions of positive/negative daily NAO indices drawn from are fitted to re-analysis. To generate a single ensemble member prediction, we do the same thing, except the persistence probabilities are first relaxed back towards $1/2$ using a scaling parameter $k$, which we refer to as the `regime fidelity' parameter. Setting $k = 0$ means the ensemble members will, on average, have persistence probabilities exactly $1/2$, while $k = 1$ means the members will, on average, have persistence probabilities matching the true ones. In this way we can generate an ensemble prediction and simulate a hindcast, where the persistence probabilities are varying each year. For technical reasons it is easier to work with log-odds rather than probabilities, which makes the implementation of this idea slightly more involved. The full details are explained in the next section.

We have also tested a method aimed at simulating the impact of (3) and (4), which we did by systematically relaxing the persistence probabilities back towards a specified climatological mean of $0.8$, thereby reducing interannual variability (whether as a result of errors in global dynamics or more complex internal dynamics). The results of the two experiments were qualitatively very similar, so we will restrict ourselves to the former. It is plausible that any bias which introduces errors in the persistence probabilities is likely to produce similar results.

\subsection{Technical formulation}
\label{sec:markov_formulation}

\begin{figure}[p]
	\centering
    \includegraphics[width=0.8\textwidth]{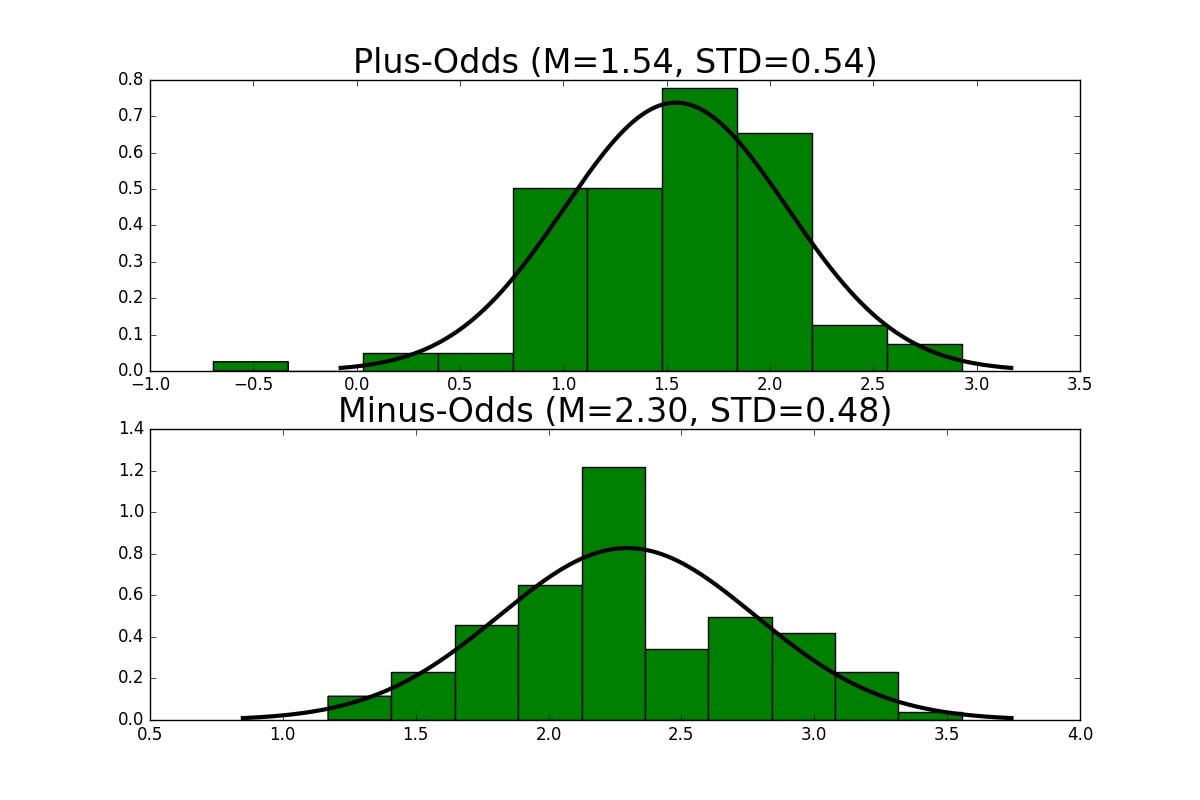}
	\caption{Histograms of log-odds of persistence probabilities of NAO+ (upper) and NAO- (lower), along with best-fit normal distributions (black curves) of both. Recall that NAO- here refers to the 3 standard regimes NAO-, Atlantic Ridge and Blocking all merged into one.}
	\label{fig:logodds_histograms}
\end{figure}

Simulating an ensemble forecast of the NAO in this framework will require the generation of randomly drawn persistence probabilities as well as the specification of the distributions $X_{+}$ and $X_{-}$. To facilitate the former, we always work with \emph{log-odds} rather than probabilities when manipulating or drawing probabilities. From an assumed distribution of probabilities $p$ (taking values between 0 and 1), a new, more Gaussian distribution $\hat{p}$ (taking values between $-\infty$ and $+\infty$) is obtained via the log-odds transformation 
\begin{equation}
\hat{p} = log(\frac{p}{1-p}). \nonumber
\end{equation}
Specific values $\hat{p}_s$ drawn from this latter distribution correspond to the probability $p_s$ from $p$ given by
\begin{equation}
p_s = \frac{1}{1+e^{-\hat{p}_s}}. \nonumber
\end{equation}
Figure \ref{fig:logodds_histograms} shows the histograms of the log-odds of persistence probabilities of NAO+ and NAO-, estimated with ERA-Interim. Note that here the NAO- regime is explicitly formed by combining, from the 4 regimes derived by K-means clustering, the 3 regimes that are not NAO+ as one. Using standard nomenclature, the true NAO- regime, along with the Atlantic Ridge and Blocking regimes, have all been combined into one regime, which we will refer to again as NAO-. The distributions appear to be fairly well approximated by normal distributions, with means (standard deviations) of 1.54 (0.54) and 2.30 (0.48) respectively. When transformed back to probabilities, these means correspond to 82\% and 91\% respectively.

\begin{figure}[p]
	\centering
    \includegraphics[width=0.8\textwidth]{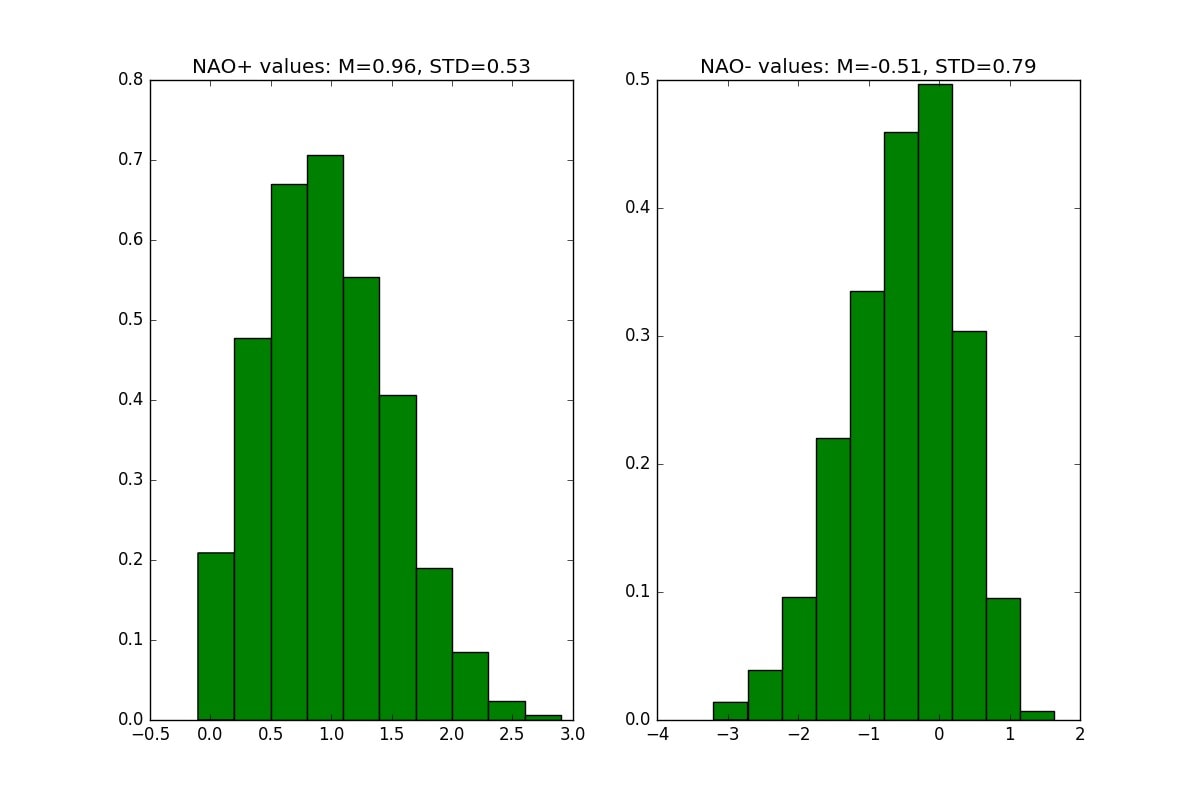}
	\caption{Projections of the NAO+ (left) and NAO- (right) regimes onto the NAO index, using ERA-Interim. Recall that NAO- here refers to the 3 standard regimes NAO-, Atlantic Ridge and Blocking all merged into one.}
	\label{fig:regime_projections}
\end{figure}

In order to estimate the distributions $X_{+}$ and $X_{-}$ we again use ERA-Interim. We simply look at the value of the NAO index\footnote{Obtained from ERA-Interim as the leading EOF of geopotential height at 500 hPa over the Euro-Atlantic region.} on each day for which the atmosphere was deemed to be in the NAO+ regime. The histogram of these numbers define the projection of the NAO+ regime onto the NAO index. Similarly we obtain the projection of the NAO- regime onto the NAO index. Figure \ref{fig:regime_projections} shows the result of this. As expected, the NAO+ projection is almost exclusively positive, while the NAO- projection is mostly negative, with some overlap onto the positive quadrant. These positive values are associated with events that really belong to either the Atlantic Ridge or Blocking regimes, which can be shown to have a more neutral projection onto the NAO index. 

The asymmetry between the distributions, along with the fact that they are both notably skewed, takes us away from the idealized picture of section \ref{sec:markov}. As we are considering an idealized situation, we will artificially rectify this discrepancy. We will firstly ignore the skewness, and assume the NAO+ projection is normally distributed, and so set $X_{+} \sim \mathcal{N}(1.0, 0.5)$. We then set $X_{-}$ to be the symmetric image, $X_{-} \sim \mathcal{N}(-1.0, 0.5)$. We verified that the three core metrics we are interested in are insensitive to this simplification.

The simulation of an ensemble forecast of a single DJF season now proceeds as follows. We start by specifying two persistence probabilities $p^{+}_{obs}$ and $p^{-}_{obs}$ for the NAO+ and NAO- regimes. These are to be interpreted as the `true' probabilities for a given DJF season. These are obtained by drawing two random log-odds numbers $L^{+}_{obs}$ and $L^{-}_{obs}$ from the fitted normal distributions of the logodds histograms from figure \ref{fig:logodds_histograms}. These are then transformed back into actual probabilities as explained earlier.

We perform the same procedure to generate an ensemble forecast, with the only exception being the choice of persistence probabilities, which we will modify according to a parameter $0 \leq k \leq 1$, referred to as the \emph{regime fidelity} factor, and a noise factor $\epsilon$, corresponding to variability between ensemble members. For every season, we now want to relax the models persistence probabilities, $p_{mod}^{+}$ and $p_{mod}^{-}$, back towards 1/2. Because $log(\frac{1/2}{1-(1/2)}) = 0$ and the log-odds function is monotonic, this is equivalent to relaxing the log-odds of these two probabilities, $L_{mod}^{+}$ and $L_{mod}^{-}$, back towards $0$. This is achieved by setting
\begin{eqnarray}
L_{mod}^{+} = k\cdot L^{+}_{obs} + \epsilon^{+}   \\
L_{mod}^{-} = k\cdot L^{-}_{obs} + \epsilon^{-} 
\end{eqnarray}
where $\epsilon^{+}$ and $\epsilon^{-}$ are random draws from a normal distribution $\mathcal{N}(0, 0.5)$. This noise factor represents the variability between ensemble members, allowing for individual members to perform better or worse by chance. The standard deviation of $0.5$ was chosen as it is approximately the standard deviation of the climatological distributions of ERA-Interim (see figure \ref{fig:logodds_histograms}). That is, we assume that the distribution of persistence probabilities across the ensemble has a spread equal to that of climatology itself. In the bimodal system considered here, this amounts to a form of model error as well, since it is assumed that the correct persistence probabilities are fixed by the ambient forcing, and so in a perfect model, every member should have the same persistence probabilities, matching those of reality. However, this `perfection' of the ensemble was considered unrealistic, and produces hindcast simulations where the ensemble has too small a variance. The climatological ensemble spread to persistence added here was considered a simple, pragmatic default assumption. The impact of this choice on the results will be discussed later. Once $L_{mod}^{+}$ and $L_{mod}^{-}$ have been obtained in this manner, they are transformed back into actual probabilities to be fed into the Markov model.

Because, as mentioned above, the logistic function applied to 0 returns 1/2, a value of $k=0$ corresponds to relaxing persistence probabilities back to $1/2$ on average (with deviations from this mean given by the $\epsilon^{+}$ and $\epsilon^{-}$ terms). When $k=1$, the ensemble members have persistence probabilities on average equal to the true probabilities. Intermediate values scale linearly between these two extremes. This source of error ($k<1$) can therefore be thought of as weak regime structure, with the potential wells of the two states being too shallow; persistence is low and therefore predictability is impaired.

Now, given a choice of the regime fidelity factor $0\leq k \leq 1$ we can generate our ensemble prediction, where for a given season, each member is the result of running a 90-day Markov chain with parameters determined as above: the persistence probabilities are therefore different in each season. Mimicking the conditions in \cite{Dunstone2016}, the hindcast length is set to 35 years and the ensemble size to 40.

\begin{figure}[p]
	\centering
    \includegraphics[width=0.8\textwidth]{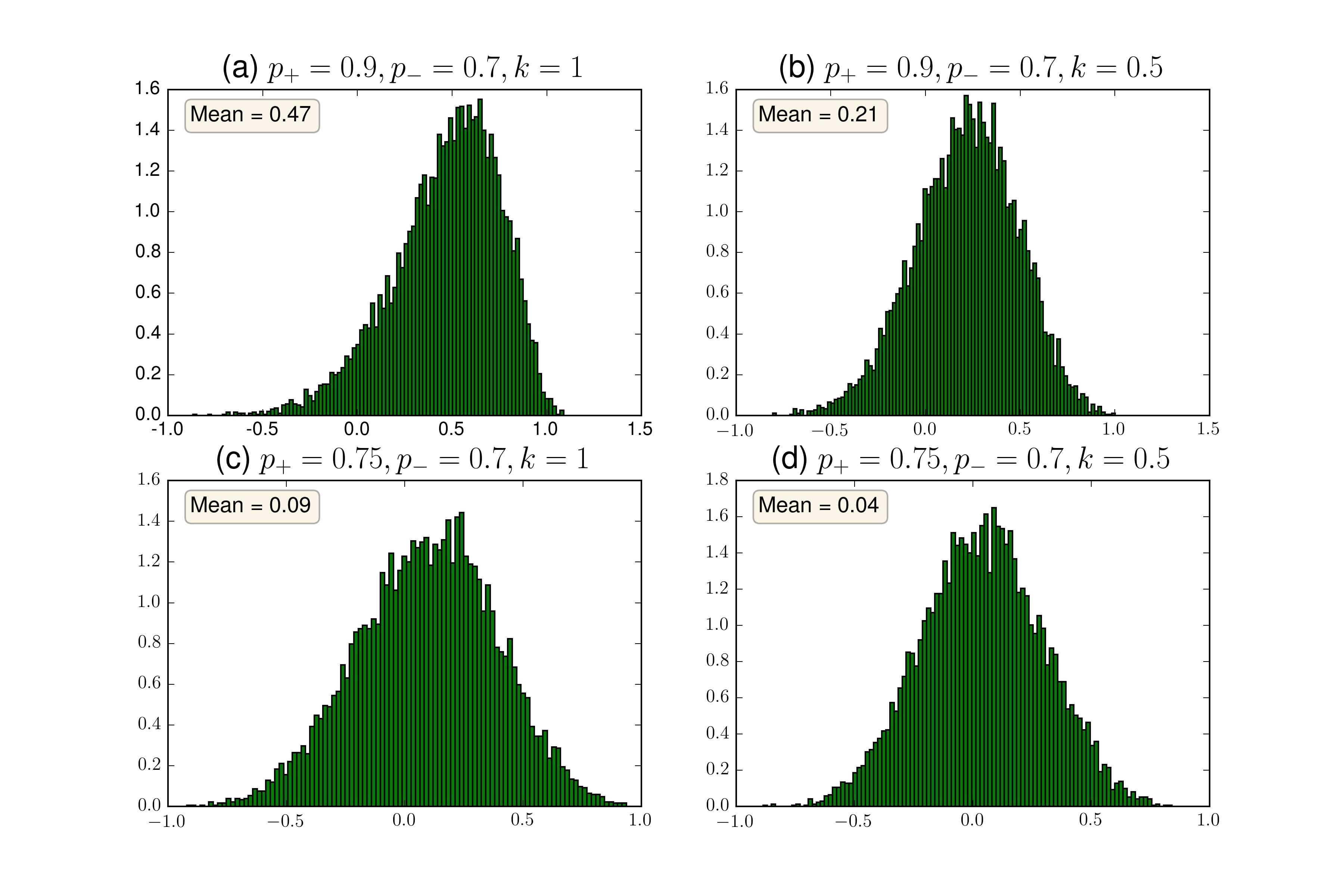}
	\caption{Histograms of seasonal mean ensembles using the Markov model. Plots (a) and (b) show a NAO+ year, with $p_{+}=0.9, p_{-}=0.7$; in (a) regime fidelity is 1 (no persistence bias), in (b) the regime fidelity is 0.5. Plots (c) and (d) show a less predictable NAO+ year, with only a weak preference; $p_{+}=0.75, p_{-}=0.7$. In (c) regime fidelity equals 1, in (d) regime fidelity equals 0.5. The histograms were produced from 10000 samples each.}
	\label{fig:markov_pdfs}
\end{figure}

Figure \ref{fig:markov_pdfs} verifies the theoretical predictions deduced in the previous section, showing seasonal mean pdfs computed with this Markov model. Plots (a) and (b) show a situation where $p_{+}=0.9$ and $p_{-}=0.7$, so the preferred state is NAO+ to a relatively high level of predictability. In (a) there is no persistence bias ($k=1$), while in (b) the regime fidelity is set to $0.5$. One can see that the perfect skill pdf is very close to being strictly positive as expected, even when sampled over just 90 days. The insertion of error weakens this, with the pdf now still being mostly positive, but admitting a larger proportion of negative values as well. As a result, the ensemble mean is much smaller in magnitude; less than half as large as the perfect skill magnitude. Plots (c) and (d) show the same but for a year with very little predictability: $p_{+}=0.75, p_{-}=0.7$, again with $k=1$ on the left and $k=0.5$ on the right. The pdfs are close to random noise with mean 0, with the $k=1$ pdf having a mean of 0.09 and the $k=0.5$ pdf having a mean of 0.04, again, less than half the magnitude of the perfect skill case.

\subsection{Correlations and RPC: dependence on regime fidelity}
\label{sec:corrs_markov}

\begin{figure}[p]
	\centering
    \includegraphics[width=0.8\textwidth]{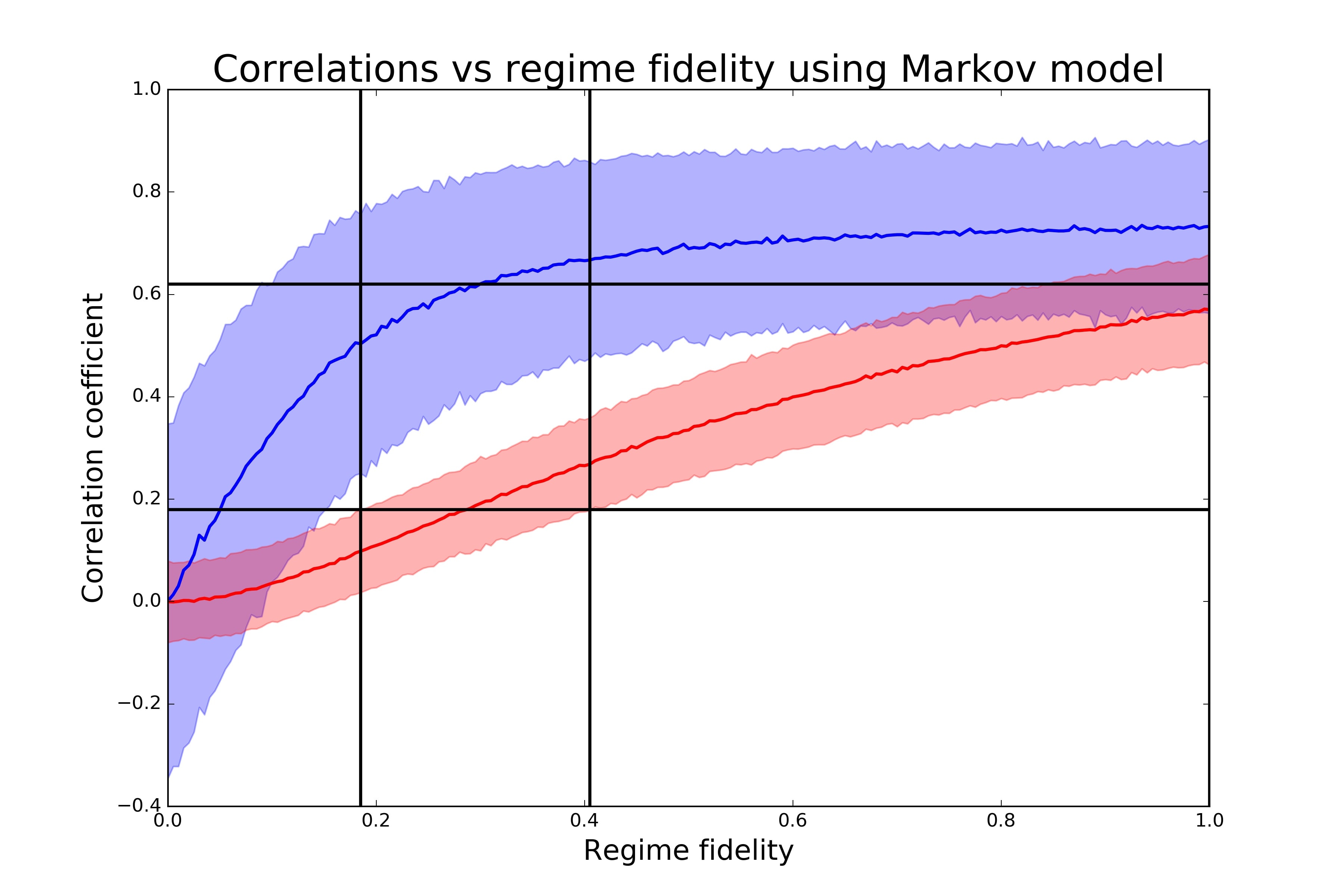}
	\caption{Correlations for the Markov model. Corr(EnsMean, Obs), the upper curve in blue, denotes the expected correlation between the ensemble mean and observations, while Corr(EnsMean, Mem), the lower curve in red, denotes the expected model predictability i.e. the average correlation between ensemble mean and individual members. Each correlation is plotted as a function of the regime fidelity parameter $k$. Shading indicates two standard deviations from the mean. Both the mean and standard deviations are computed, for each $k$, over a sample of size 1000. Horizontal lines indicate the values seen in DePreSys3 of 0.62 and 0.18, while vertical lines isolate a critical range for which these values are plausibly obtained (see main text).}
	\label{fig:markov_corrs}
\end{figure}

Figure \ref{fig:markov_corrs} shows the actual and model predictability for the Markov model as a function of regime fidelity $k$: shading gives an indication of sampling variability, given as two standard deviations from the mean values. For each value of regime fidelity factor $k$, 1000 simulations of a 40-member ensemble hindcast were performed to generate these statistics. Horizontal lines indicate the DePreSys3 values, i.e. actual predictability 0.62 and model predictability 0.18. Where these intersect the thick blue and red lines indicates where obtaining these values is expected. By intersecting the horizontal lines with the shading, we can isolate, for each value, a region of regime fidelity values for which obtaining either DePreSys3 value lies within the confidence interval. This allows us to isolate a critical region of skill values for which both DePreSys3 values can be plausibly obtained: this region is marked with vertical lines and corresponds approximately to $0.18 < k < 0.41$.

For example, when $k=0.3$, the expected values are
\begin{eqnarray}
Corr(EnsMean, Obs) = 0.63 \, (0.43, 0.83)  \\
Corr(EnsMean, Mem) = 0.19 \, (0.11, 0.27)  \\
RPC_{est} = 2.28 \, (1.46, 3.11) \nonumber
\end{eqnarray}
where the bracketed values are the lower and upper bounds of the 95\% confidence interval. The expected values are virtually the same as those obtained by DePreSys3.

\begin{figure}[p]
	\centering
	\includegraphics[width=0.8\textwidth]{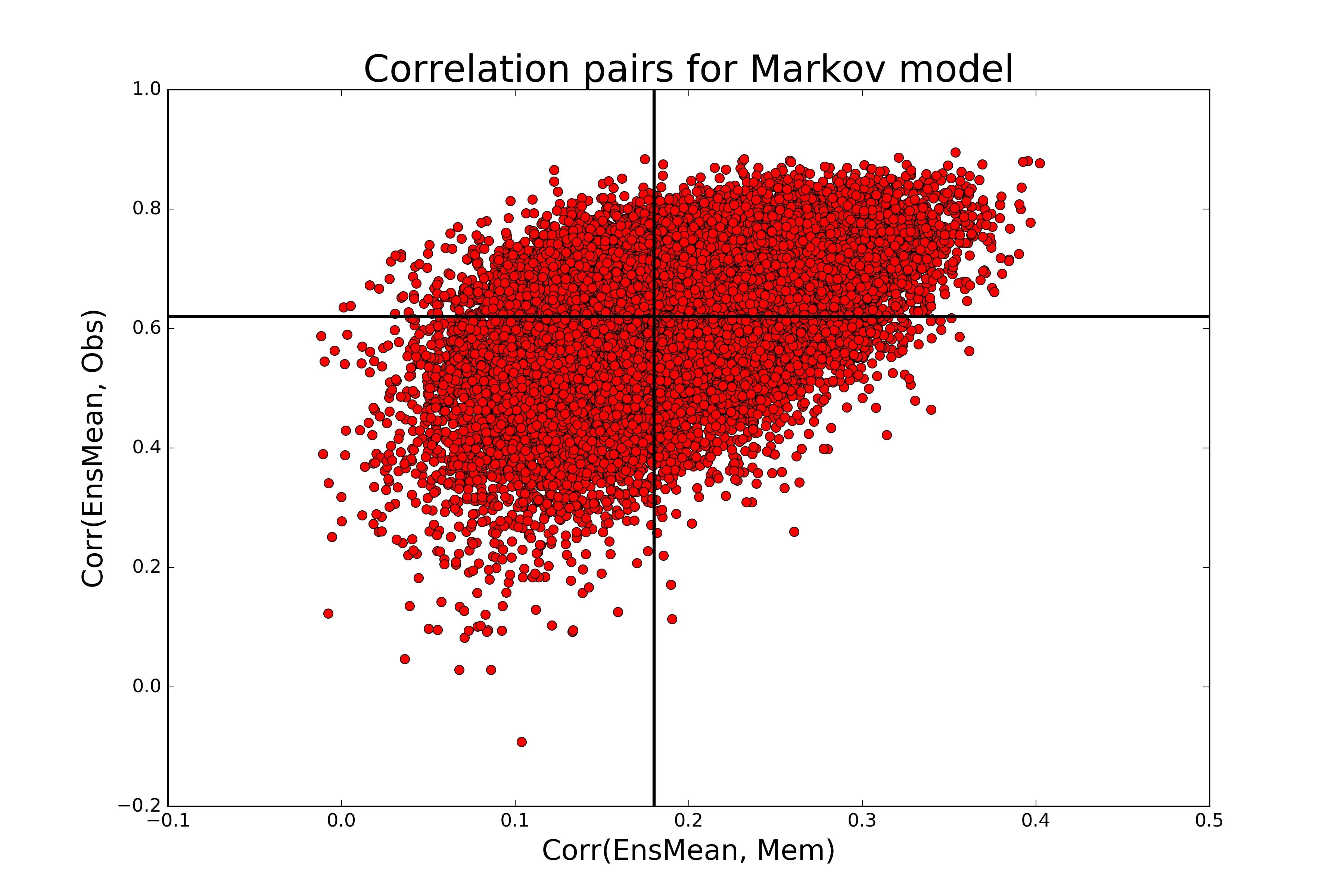}
	\caption{Actual pairs of correlations obtained from simulations using the Markov model. For a chosen value of regime fidelity $k$ in the critical range (see main text), we ran 1000 simulations, and for each such simulation computed a single pair. Fifty choices of $k$ were made, evenly sampled within the critical range, so the figure shows a total of 50000 such pairs. The intersection of the horizontal and vertical line marks the pair (0.18, 0.62) observed in DePreSys3.}
	\label{fig:markov_corrpairs_critrange}
\end{figure}

\begin{figure}[p]
	\centering
	\includegraphics[width=0.8\textwidth]{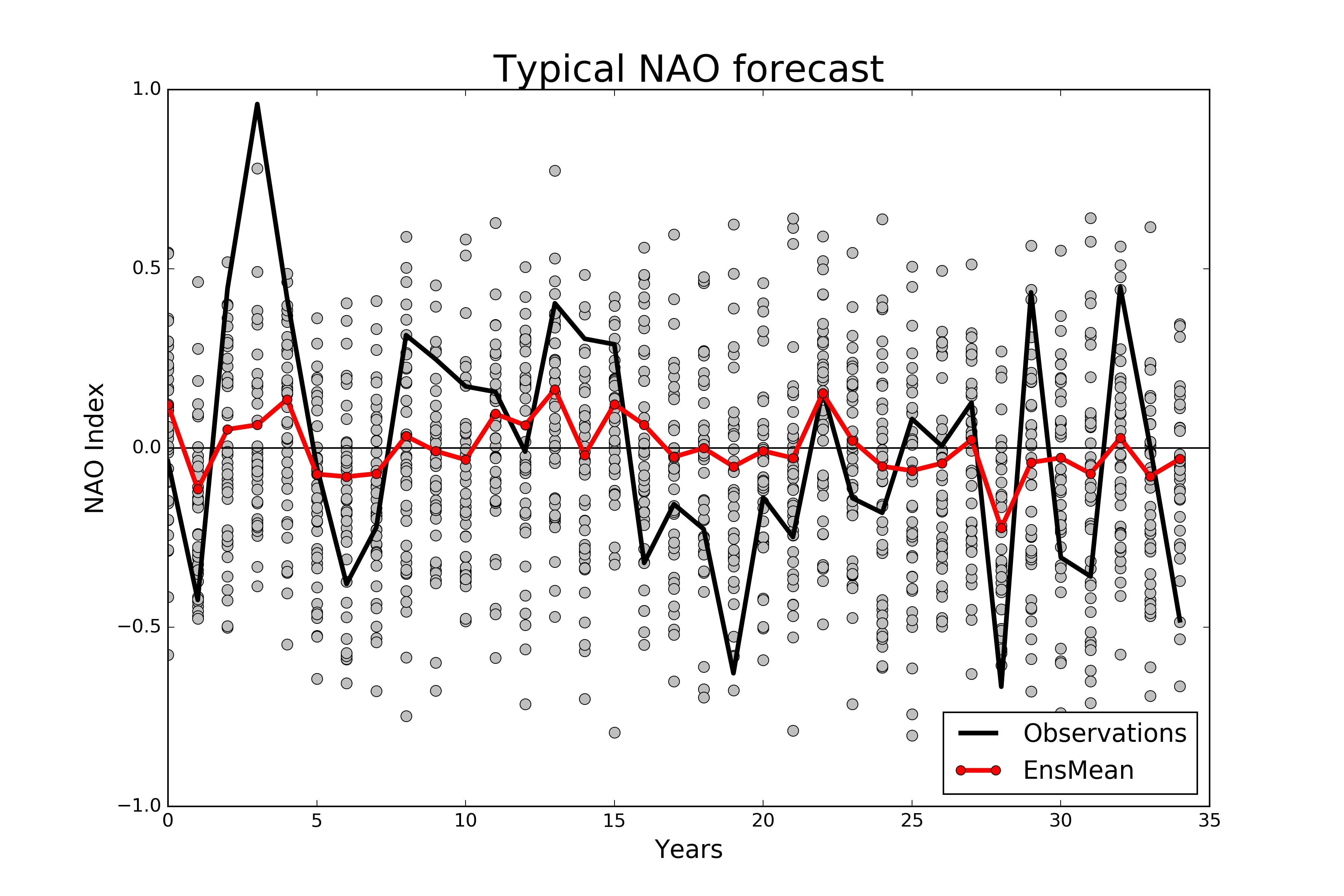}
	\caption{A typical example of a hindcast simulation using the Markov model with regime fidelity $k=0.3$. Individual ensemble members are displayed as grey dots, with the ensemble mean in red (dotted). In black is the randomly generated `observations' for this simulation. In this hindcast, actual predictability was $0.55$, model predictability was $0.14$, and $RPC_{est}$ was $2.31$.}
	\label{fig:typical_forecast_unscaled}
\end{figure}

\begin{figure}[p]
	\centering
	\includegraphics[width=0.8\textwidth]{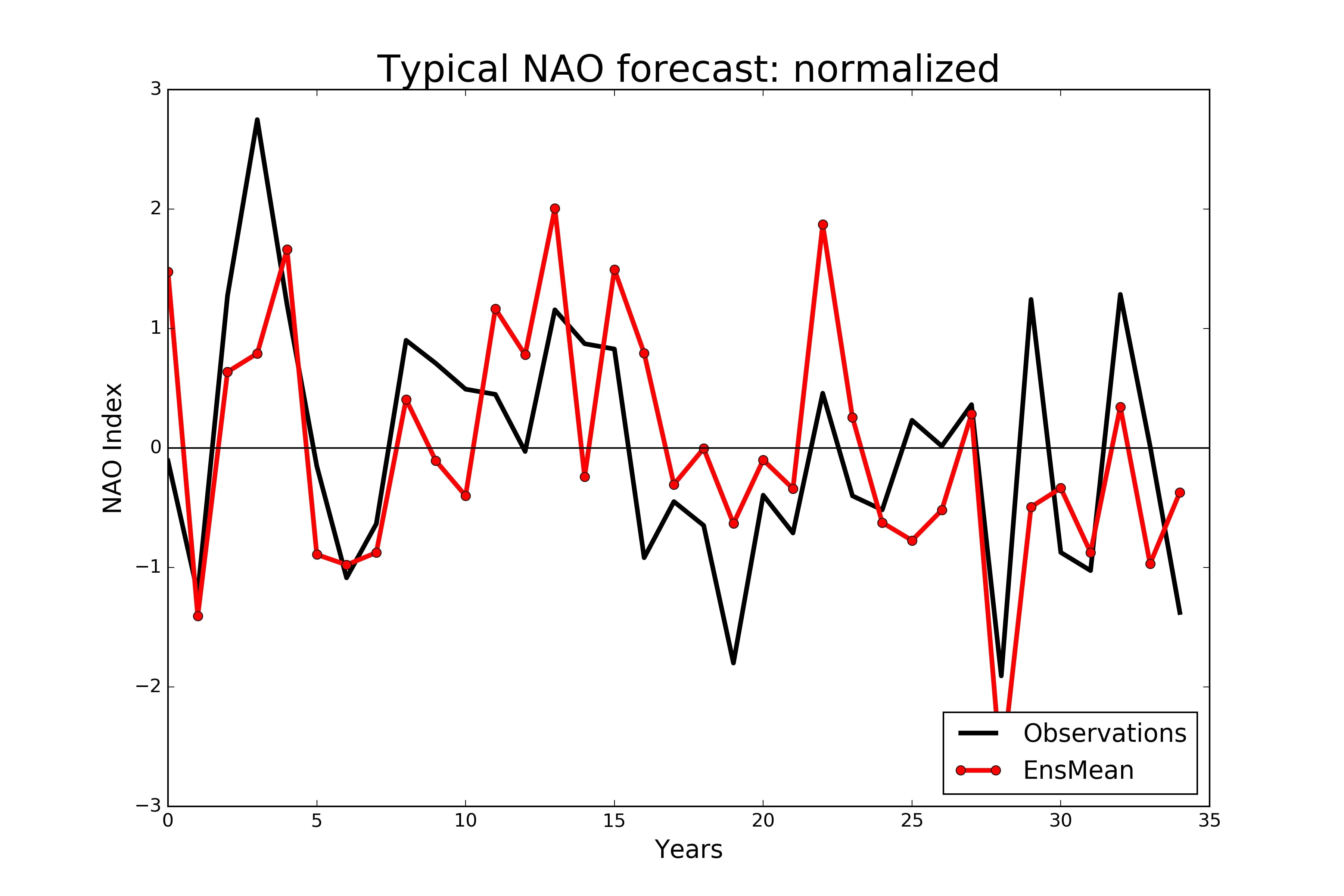}
	\caption{The same forecast as in figure \ref{fig:typical_forecast_unscaled}, but now ensemble mean (red dotted) and reality (black) have been normalized to have standard deviation 1.}
	\label{fig:typical_forecast_scaled}
\end{figure}

Because the two correlations are not independent, and always come as a \emph{pair} associated to a specific simulation, it is possible that while each individual number is easy to obtain, the pair itself is unlikely. To estimate the true joint probability, we have in figure \ref{fig:markov_corrpairs_critrange} explicitly plotted around 50000 such pairs, obtained from simulations where we restricted ourselves to values of $k$ in the critical range only. It can be seen that the DePreSys3 pair lies at the centre of the cluster, indicating that obtaining this exact pair is very plausible with this model. In figure \ref{fig:typical_forecast_unscaled} we have plotted a typical example of a hindcast simulation with $k=0.3$. Figure \ref{fig:typical_forecast_scaled} shows the same, except we have normalized the ensemble mean and observations to have standard deviation 1. In this example, the three main metrics were $0.55, 0.14$ and $2.31$. These figures should be compared to the equivalent figures 1a and 2a in \cite{Dunstone2016}.

Observe that in figure \ref{fig:markov_corrs}, actual predictability increased much more rapidly with $k$ than model predictability does. This can be understood conceptually in the following way. The seasonal mean is to a large extent captured by the preferred regime state, which is in turn determined, via equation (9), by the \emph{difference} between the two persistence probabilities. The ensemble mean will, for a sufficiently large ensemble size, have persistence probabilities determined by equations (11) and (12), . As such, both reality and the ensemble mean will, for a large ensemble, have the same preferred state in each season and correlate strongly with each other. However, many individual ensemble members will firstly, due to the internal variability, have sufficiently different persistence probabilities to ensure they have a different preferred regime. Secondly, other members, while still retaining the preferred regime of reality, will end up with the probabilities for both regime being closer together (both being close to $1/2$): in this setting of reduced predictability such members may well end up realizing the wrong preferred regime state simply due to sampling variability. This would correspond, in figure \ref{fig:markov_pdfs}(c), to a member ending up in the negative quadrant. The ensemble mean will correlate poorly with all such members. In total therefore, the ensemble mean will be expected to correlate more consistently strongly with reality than with individual ensemble members.

\begin{figure}[p]
	\centering
    \includegraphics[width=0.8\textwidth]{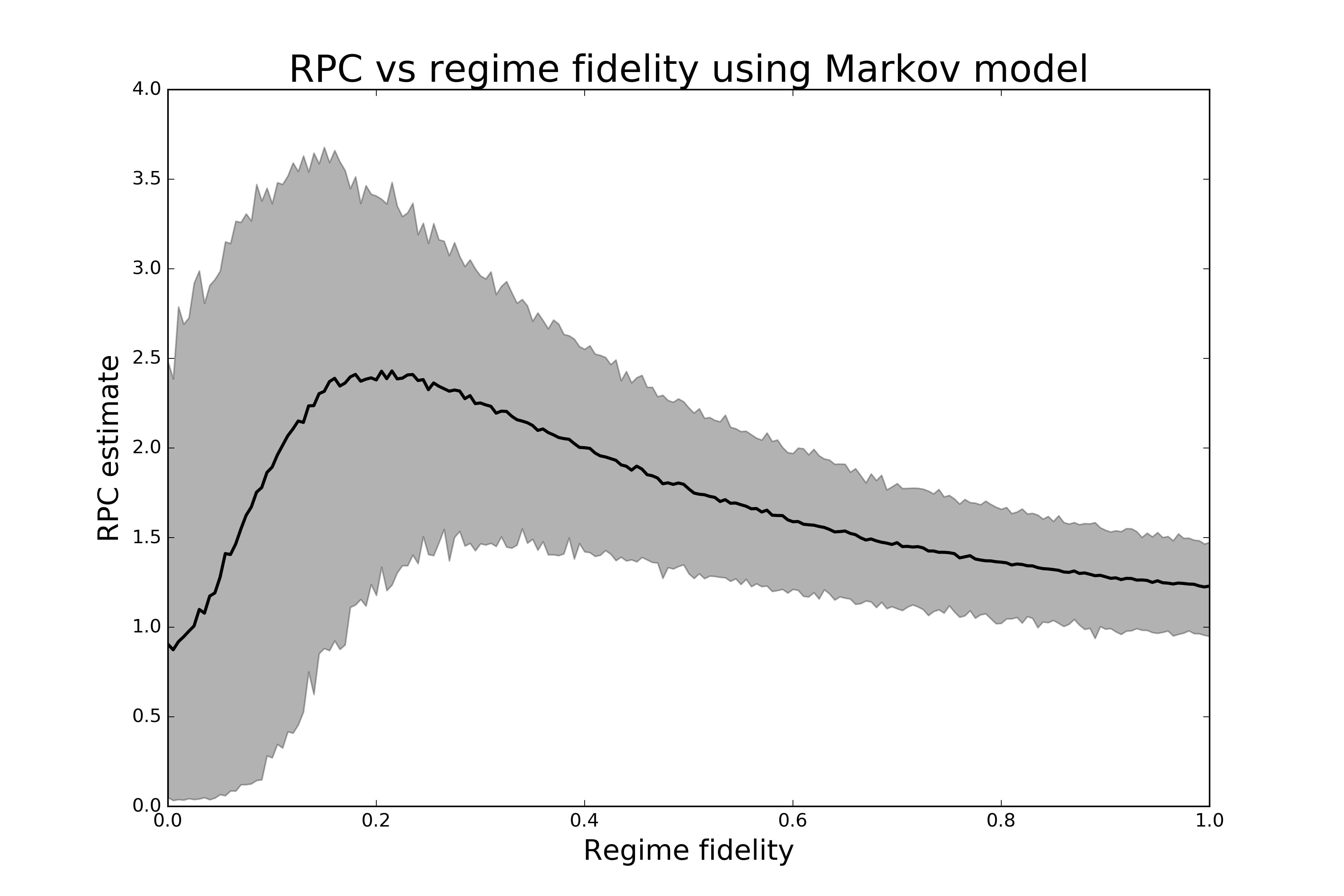}
	\caption{$RPC_{est}$ as a function of regime fidelity $k$ (see main text) using the Markov model. The thick line shows the expected value for the given choice of skill, while shading denotes the 95\% confidence interval. For each skill value, 1000 simulations were used to generate these statistics.}
	\label{fig:rpc_markov}
\end{figure}

This is naturally reflected also in the $RPC$. Figure \ref{fig:rpc_markov} shows the $RPC_{est}$ as a function of regime fidelity $k$. With maximal regime fidelity ($k=1$), the expected $RPC_{est}$ is around 1.3. The fact that it isn't exactly $1$ is due to the internal variability between ensemble members. If this were eliminated, so that all ensemble members have the exact same persistence probabilities, then the expected $RPC_{est}$ with $k=1$ is $1$. Since around 75\% of the discrepancy between the DePreSys3 value of 2.3 and the desired value of 1 is eliminated by improving the regime fidelity, we conclude that the dominant cause for obtaining $RPC_{est} > 1$ in this system is due to insufficient regime fidelity. 

\begin{figure}[p]
	\centering
    \includegraphics[width=0.8\textwidth]{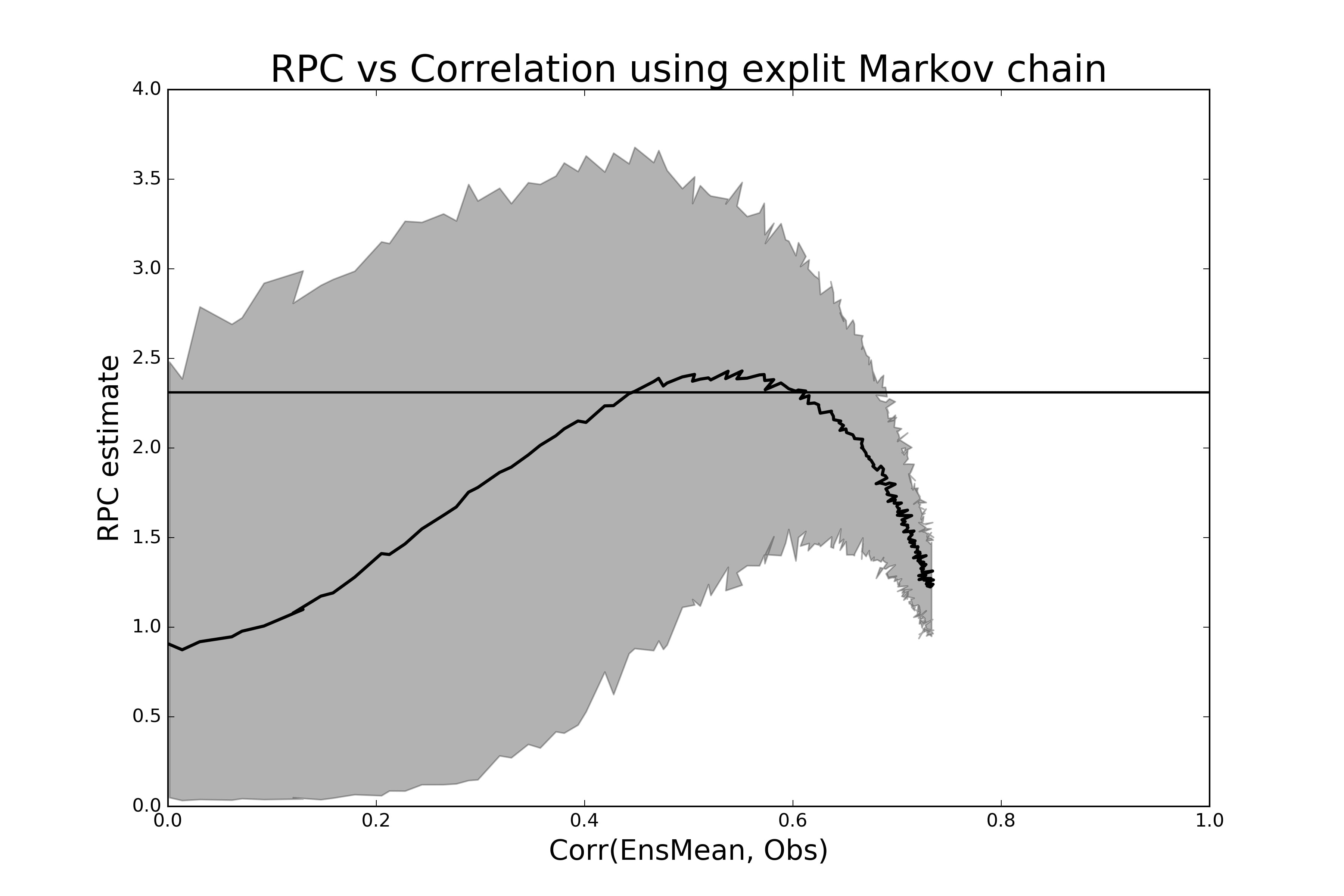}
	\caption{$RPC_{est}$ as a function of ensemble mean correlation, using the Markov model. The thick line shows the expected value for the given level of correlation, while shading denotes the 95\% confidence interval. For each correlation coefficient, 1000 simulations were used to generate these statistics. The horizontal line marks the value of $2.31$ obtained by DePreSys3.}
	\label{fig:rpc_v_corrs_markov}
\end{figure}

Observe that in figure \ref{fig:rpc_markov}, the change in slope at around $k=0.2$ is due to the divergence of $RPC_{est}$ from $RPC_{true}$. For $k<0.2$, the model error has grown large enough to ensure that an ensemble mean of just 40 members no longer accurately represents reality. Figure \ref{fig:rpc_v_corrs_markov} shows $RPC_{est}$ as a function of actual predictability. It can be seen that this divergence in the estimated $RPC$ from true $RPC$ begins to take effect for correlations of $0.4$ and lower. This suggests that $RPC$ estimates in situations where the ensemble mean correlation is less than 0.4 are unlikely to be meaningful; this is also reflected in the large errorbars in this region. Of particular importance here is that, as explained in section \ref{sec:caution}, the expected $RPC_{est}$ at $k=0$ here is around $0.85$. Due to the large errorbars, it is therefore likely that, when estimating the $RPC$ of a model with insufficient skill, one may compute $RPC_{est} \approx 1$ even when $RPC_{true} \gg 1$.

Our conclusion is that this idealized Markov model captures well the observed $RPC$ behaviour of DePreSys3. Importantly, it does so even though the models internal level of noise, on \emph{daily} timescales, is exactly identical to that of observations, both being measured by the standard deviation of $X_{+}$. The apparently unrealistic amount of noise on \emph{seasonal} timescales, as measured with the signal+noise model in \cite{Siegert2016}, is in this context a purely emergent feature of model error. In other words, a weak signal with realistic noise on \emph{daily} timescales can end up looking like a weak signal with a large amount of noise on \emph{seasonal} timescales, when filtered through a regime system.

Finally, we note that a regime fidelity factor $k=0.3$ corresponds roughly to the ensemble mean underestimating the true seasonal persistence probabilities (relative to ERA-Interim) by a factor of $0.8$ on average. For comparison, the GloSea5 data considered in section \ref{sec:error_sources} shows an underestimate by a factor closer to $0.9$. While a fidelity factor $k=0.41$, the highest possible factor still exhibiting the desired behaviour, brings this factor of $0.9$ within the range of sampling variability, this is likely indicative of the two-state system being too crude an approximation.

\subsection{Correlations and RPC: dependence on ensemble size}
\label{sec:corrs_enssize_markov}

\begin{figure}[p]
	\centering
    \includegraphics[width=0.8\textwidth]{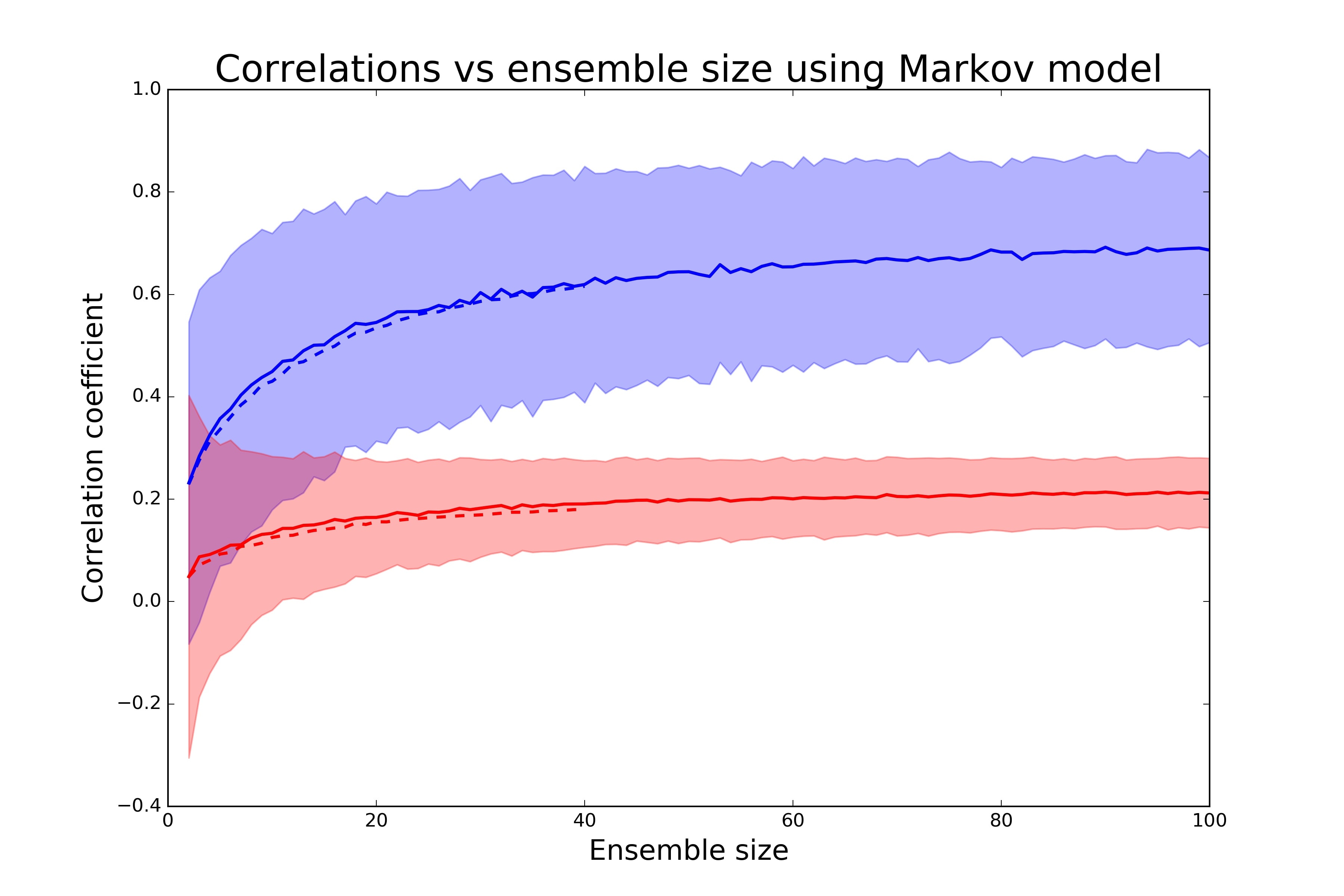}
	\caption{Correlations as a function of ensemble size. Actual predictability, the upper curve in blue, shows the expected correlation between the ensemble mean and observations, while model predictability, the lower curve in red, shows the average correlation between ensemble mean and individual members. Thick lines show the expected value for the given ensemble size, while shading denotes the 95\% confidence interval; for each choice of ensemble size, 1000 simulations were performed to generate the statistics. Regime fidelity was fixed at $k=0.3$. The stipled lines show estimates of the same quantities from DePreSys3. }
	\label{fig:markov_enssize_corrs}
\end{figure}

Figures \ref{fig:markov_enssize_corrs} show how correlations grow as a function of ensemble size, where the regime fidelity factor $k$ has been fixed at $0.3$. The thick coloured lines show the expected actual and model predictability, while shading indicates the 95\% confidence interval. Stipled lines show estimates of the same quantities as computed from raw DePreSys3 data. We see again that the Markov model captures the relationship very well.\footnote{We remind the reader that the linear model in \cite{Siegert2016} is also able to capture this well, and so the Markov model's good agreement here is not a unique feature.}.

In this model of the NAO, the benefit of increasing the ensemble size begins to saturate after around 80 members. With 100 members, the expected ensemble mean correlation is around $0.7 \pm 0.2$. This is roughly the same as what can be obtained by setting the regime fidelity to $1$, suggesting that one cannot expect to gain more skill by increasing the ensemble size further. By maximizing the regime fidelity, removing internal variability between members (so all members share the same persistence probabilities) and setting the ensemble size to 100, the expected correlation can be increased to around $0.75 \pm 0.22$, the maximal level of skill obtainable here. Indeed, the expected correlation was found to be the same when increasing to 1000 ensemble members.

\begin{figure}[p]
	\centering
    \includegraphics[width=0.8\textwidth]{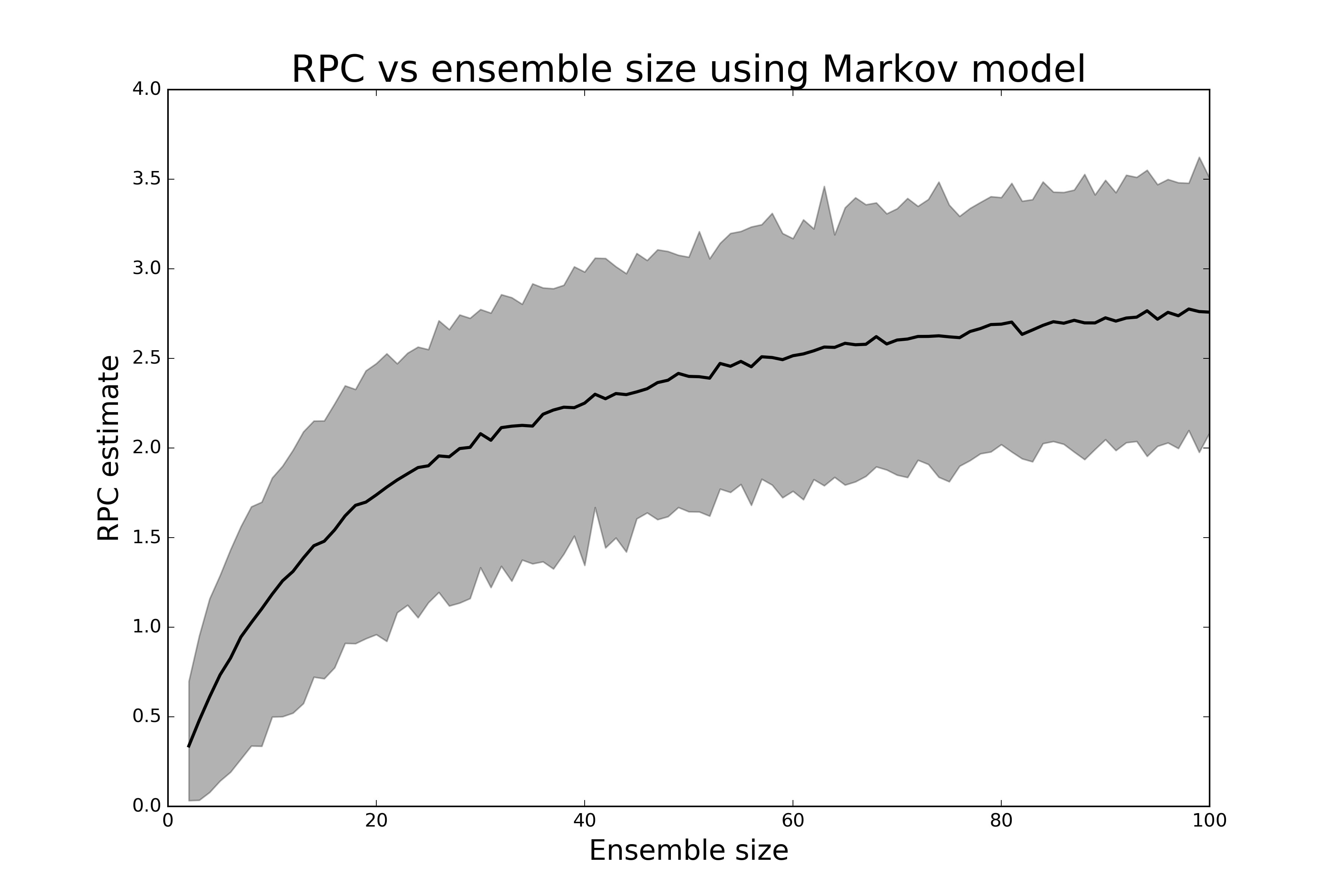}
	\caption{$RPC_{est}$ as a function of ensemble size, using the Markov model. The thick line shows the expected value for the given choice of skill, while shading denotes the 95\% confidence interval. For each skill value, 1000 simulations were used to generate these statistics. The regime fidelity was fixed at $k=0.3$.}
	\label{fig:rpc_v_enssize_markov}
\end{figure}

Figure \ref{fig:rpc_v_enssize_markov} shows $RPC_{est}$ versus ensemble size. As ensemble size increases, the ensemble mean becomes a better approximation of reality, and $RPC_{est}$ becomes a better estimate of $RPC_{true}$. Since $RPC_{true} \geq RPC_{est}$, this explains why $RPC_{est}$ increases with ensemble size. With 100 members, $RPC_{est}$ is approximately $2.7$. The flattening of the growth of $RPC_{est}$ in figure \ref{fig:rpc_v_enssize_markov} suggests that this is beginning to approach the true value $RPC_{true}$. With 1000 members, $RPC_{est} = 3.2$ with a 95\% confidence interval of $[2.48, 3.96]$, implying that the true $RPC$ in this bimodal system is greater than $2.5$ with high confidence.

A final point of interest here is that the scaling of these metrics with ensemble size provides one conceptually easy way to distinguish between different statistical models. For example, both the signal+noise model of \cite{Siegert2016}, and our Markov model here, were fitted to a specific ensemble size, and therefore will be in good agreement with each other for this particular size. However, the manner in which they then scale when the ensemble size varies is different for the two models. In principle, this offers a method for distinguishing between the two models, though due to sampling uncertainty, the ensemble size would need to be on the order of 1000 to robustly detect the difference.

\section{Over or under-confident: RPC vs RMS error/spread}

One way to interpret a model with $RPC>1$ is that the model is `under-confident', in the sense that individual members deviate from the ensemble mean more than is appropriate given the true level of predictability of the system. Another widely used method to assess if a model has an appropriate level of noise relative to its skill is to compare the average root-mean-square error of the ensemble mean to the total ensemble spread (\cite{Johnson2009}). A derivation of why these should be expected to match for a perfectly calibrated ensemble can be found e.g. in \cite{Fortin2014}, which also includes precise definitions for the terms (in particular the correct definition of ensemble spread). If the error exceeds the spread, this could be understood as the model being `over-confident' (the ensemble pdf tends to be sharply centered at the wrong location), while if the spread exceeds the error, this is indicative of the model being `under-confident' (the ensemble pdf is centered in roughly the correct place but is much wider than that of reality).

\begin{figure}[p]
	\centering
    \includegraphics[width=0.8\textwidth]{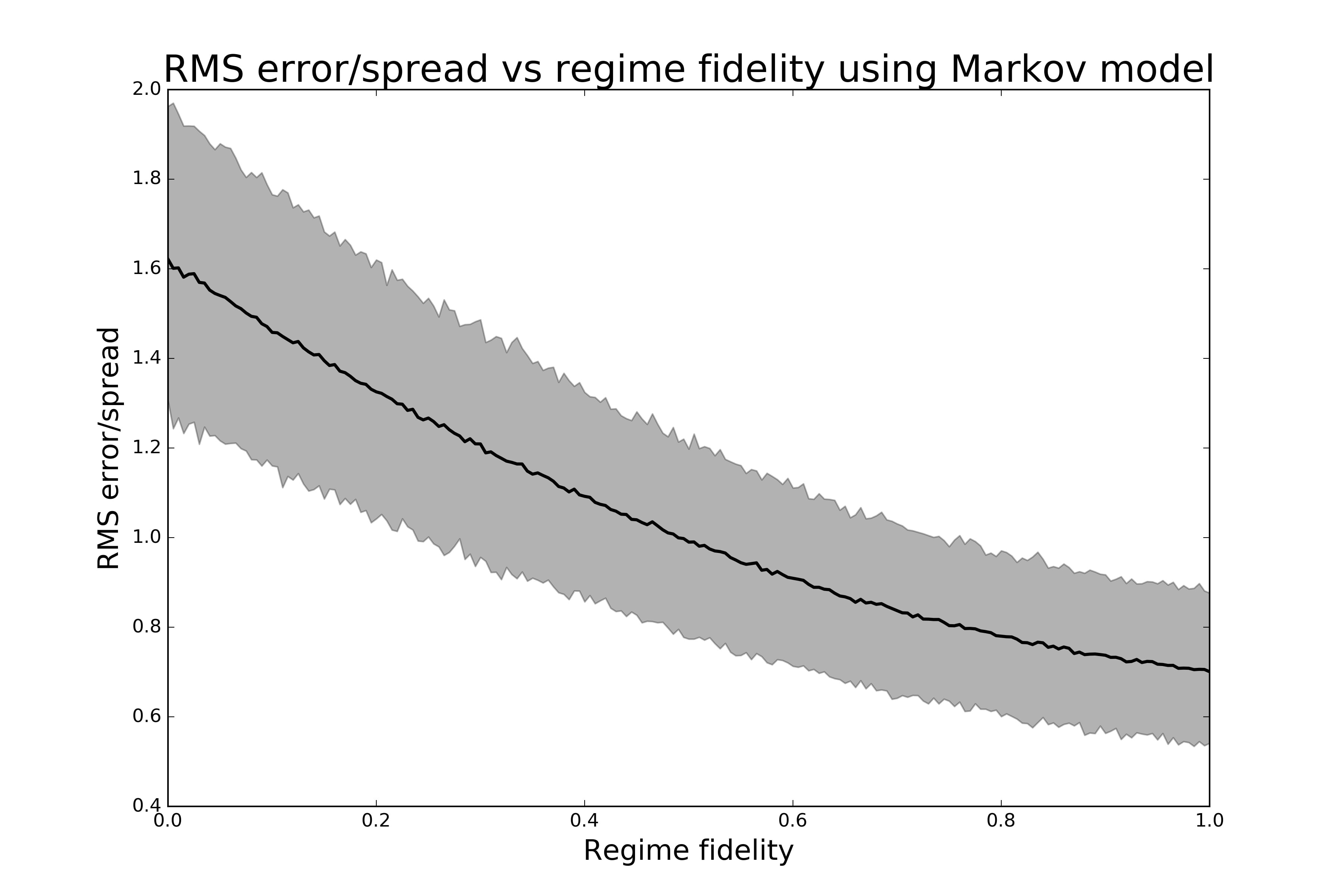}
	\caption{Root-mean-square ensemble mean error divided by ensemble spread for the Markov model, for different values of regime fidelity $k$. Shading indicates the 95\% confidence interval and the solid line the mean. For each $k$, 1000 simulations were run to obtain these statistics.}
	\label{fig:error_spread}
\end{figure}

Figure \ref{fig:error_spread} shows the RMS error over spread metric as a function of regime fidelity. It can be seen that while $RPC_{est}$ is essentially always expected to be greater than $1$, the error-spread metric changes notably as regime fidelity increases. For lower levels of regime fidelity, this metric is expected to be greater than $1$, indicative of over-confidence, while the opposite is true for high levels of regime fidelity. In general therefore one cannot expect that $RPC$ and error-spread metrics agree on whether a model is over or under-confident. By tweaking the parameters of the model one can easily obtain even more extreme examples. As expected, one can verify that when the model is statistically perfect ($k=1$ and no variability between ensemble members), one finds that both $RPC_{est}$ and RMS error/spread equal 1.

The scaling of error-spread with regime fidelity can be understood as follows. Due to equation (10), the variance of the ensemble is independent of the persistence probabilities, and is therefore going to be, on average, the same for every season. Therefore the decrease in error-spread is simply due to the ensemble mean error decreasing as regime fidelity increases. 

If one computes these numbers for the DePreSys3 hindcasts, one finds that the ratio of root-mean-square error to ensemble spread is $0.87$, indicative of a very small amount of model under-confidence. Since the overall ensemble spread matches that of observations well (see e.g. \cite{Dunstone2016}), this is due to the RMS error being slightly too small. With a regime fidelity of $0.3$, the Markov model predicts this number to be around $1.2$, suggesting that the Markov model has slightly too small a variance. For regime fidelity parameters closer to $0.4$, the upper end of the critical range, one can however obtain $0.87$ within the 95\% confidence interval of the Markov model.

An important observation here\footnote{We thank an anonymous reviewer for pointing this out.} is that in situations where the magnitude of the ensemble mean is small, the RMS error will, by virtue of its definition, typically be close to the ensemble spread, and so the ratio will be close to 1. Therefore, depending on sample size, this metric may not easily detect any problems. On the other hand, the $RPC_{est}$ metric, by virtue of the fact that correlations are insensitive to magnitude, will not have this short-coming. However, even this observation is subject to its own subtleties, visible by reference to figures \ref{fig:error_spread} and \ref{fig:rpc_markov}. For example, when regime fidelity is $0.1$, the RMS error divided by spread is robustly greater than $1$, despite the fact that the signal in this situation is tiny. At the same time, the $RPC_{est}$ metric cannot be clearly distinguished from $1$, due to the poor limiting behaviour of the metric at very low levels of model skill. This again reinforces the point that these metrics, by measuring subtly different things will behave differently for imperfect models, making the attachment of meta-level descriptors (e.g. `over/under-confident') tricky.

Finally, we note that it is the RMS error divided by spread metric which is most influenced by the added internal variability between persistence probabilities of individual ensemble members (the $\epsilon$'s in equations (11) and (12)). Removing this variability entirely results in this metric being notably too large, due to a reduction in ensemble spread. This is due to the fact that because of equation (9), if all ensemble members have persistence probabilities with the correct relative magnitude, the ensemble members as well as the ensemble mean will tend to have the correct sign (unless the difference between the two probabilities is small). This is in contrast to what is observed in \cite{Dunstone2016}, where ensemble members tend to inhabit both positive and negative values fairly evenly.

\section{Conclusion}
\label{sec:conclusion}

We constructed and analysed a statistical model aimed at capturing NAO predictability within an idealized bimodal regime setting. In this paradigm, predictability is provided by the models ability to faithfully represent the seasonally varying persistence probabilities between the different states, which in turn determine the preferred state in a given season. By fitting a single `regime fidelity' parameter $k$, representing the GCMs ability to represent persistence adequately, we were able to use this model to accurately capture both the high level of skill and the large $RPC_{est}$ values observed in the \cite{Dunstone2016}. With this fitted parameter kept fixed, we showed that the scaling of these metrics with ensemble size also matches that of \cite{Dunstone2016} extremely closely. This suggests that the high skill of DePreSys3 together with its large $RPC_{est}$ value could be because, while being able to detect changes in seasonal persistence probabilities well, it never the less retains systematic biases leading to a consistent under-estimation of persistence. Such model bias could, in this setting, be from a number of sources, including weak regime structure, poor propagation of external forcing/teleconnections, or generic errors in the initial conditions. As models are well known to have systematic problems with producing robust regimes compared to re-analysis, we speculate that alleviating this may be the most immediately accessible route towards even higher levels of NAO predictability and a better signal-to-noise ratio. While there is some evidence that both increased resolution and the implementation of stochastic physics schemes can alleviate such errors (see \cite{Dawson2015}), it is clear that there is still much work to be done in this area.

We emphasize to finish that the model tested is highly idealized, using a simplified bimodal setting. Many studies suggest there may be up to 4 regimes, where the model may be expected to have differing levels of skill and bias for all the associated 16 transition probabilities. In particular, small biases in multiple transitions may compound to produce large errors. A systematic study of this full regime system and the ability of models to capture this structure is a clear next step in illuminating this further, which the authors hope to address in future work. The results of this paper therefore should be understood primarily as a proof-of-concept result, demonstrating that the frequently observed model bias of weak regime persistence may be a crucial component to understanding NAO predictability and the `signal-to-noise paradox'.


\acks
K.S. acknowledges funding from the European Commission under Grant Agreement 641727 of the Horizon 2020 research programme. T.N.P acknowledges funding by European Research Council grant number 291406. We are also particularly grateful to Nick Dunstone et al. who graciously offered the DePreSys3 hindcast data for our analysis. Conversations with David Stephenson and Fenwick Cooper were also of great value.

\bibliographystyle{wileyqj}           
\bibliography{nao_references}{}


\section{Appendix A}

We sketch how to derive the average proportion of time spent in each state of a bimodal Markov chain with persistence probabilities $\alpha$ and $\beta$, as given in section \ref{sec:markov}. Because Markov chains are usually discussed in the literature in terms of transition probabilities, we define $a = 1 - \alpha$ and $b = 1-\beta$. Therefore we are dealing with a 2-state Markov chain with transition matrix
$$P = 
\begin{pmatrix}
a & 1-a \\
b & 1-b
\end{pmatrix}
$$
Now recall, for example by combining theorem 1.7.7 and theorem 1.10.2 from \cite{Norris1997}, that in this  setting, $P$ has a unique invariant distribution, i.e. a vector $(\pi_1, \pi_2)$ such that
$$
\begin{pmatrix}
\pi_1 & \pi_2
\end{pmatrix}
P = 
\begin{pmatrix}
\pi_1 \\
\pi_2
\end{pmatrix}
$$
where $\pi_1 + \pi_2 = 1$, and, furthermore, $\pi_1$ and $\pi_2$ are precisely the average proportion of time spent in the two states. This is a system of linear equations which, due to the constraint $\pi_1 + \pi_2 = 1$, has a unique solution, which can readily be deduced with basic linear algebra. Doing so and expressing the final answer in terms of persistence probabilities gives the expressions found in \ref{sec:markov}.


\end{document}